\documentclass{article}
\usepackage{latexsym}

\newcommand{\ga}{\alpha} \newcommand{\gb}{\beta}
\newcommand{\gc}{\gamma} 
\newcommand{\gre}{\epsilon}

\newcommand{\sbs}{\subseteq} 
 \newcommand{\lto}{\rightarrow}

 \newcommand{\cl}{{\mathcal L}}

\newcommand{\cf}{{\mathcal F}}

\newcommand{\vd}{\vdash}
\newcommand{\Cn}{{\hbox{\rm Cn}}}
\newcommand{\Coh}{{\hbox{\rm Coh}}}

\newcommand{\ev}{\makebox[1.3em]{$\rule{0.1mm}{2.5mm}\hspace{-1.35mm}
\sim$}}
\newcommand{\notev}{\makebox[1.3em]{$\rule{0.1mm}{2.5mm}\hspace{-1.35mm}
\not\sim$}}
\newcommand{\sev}{\makebox[1.8em]{$\rule{0.1mm}{2.5mm}\hspace{-1.4mm}\sim_\leq$}}
\newcommand{\notsev}{\makebox[1.8em]{$\rule{0.1mm}{2.5mm}\hspace{-1.4mm}\not
\sim_\leq$}}
\newcommand{\gmev}{\makebox[2.2em]{$\rule{0.1mm}{2.5mm}\hspace{-1.4mm}
\sim_{\hbox{\scriptsize  GM}}$}}

\newtheorem{thm}{\bf Theorem}
\newtheorem{lem}[thm]{\bf Lemma}
\newtheorem{prop}[thm]{\bf Proposition}
\newtheorem{cor}[thm]{\bf Corollary}

\newenvironment{defn}{\par\medskip\addtocounter{thm}{1}%
  \noindent{\bf Definition \arabic{thm}}\quad}{\medskip}

\newenvironment{exmp}{\par\medskip\addtocounter{thm}{1}%
  \noindent{\bf Example \arabic{thm}}\quad}{\medskip}

\newcommand{\qed}{\vrule height5pt width3pt depth0pt}
\newenvironment{pf}{\noindent {\it Proof.}}{{\nobreak\hfill \qed \par \medbreak}}

\title{To Preference via Entrenchment}

\date{January 12th, 1998}

\author{Konstantinos Georgatos\\
Graduate Program in Logic and Algorithms\\
 University of Athens\\ Panepistimioupolis 157 71, Greece}

\begin{document}
\maketitle

\begin{abstract}
We introduce  a simple generalization of G\"ardenfors
and Makinson's  epistemic  entrenchment called partial entrenchment. We show that preferential inference can be generated as the sceptical counterpart of an inference mechanism defined directly on partial entrenchment.
\end{abstract}

\section{Introduction}
\label{sec:intro}

Preference is an important concept in knowledge representation.
Whenever we aim to design a framework that does not depend solely on logical
considerations, a possible way to incorporate extralogical information
is to treat it as  preference.
Preference is subjective. Yet, preference is not based on a beyond
analysis personal taste. If that was the case, it would have been
pointless to seek a logic for preference.

Preference is based on available information, both implicit
(facts we learned and believed) and
explicit (facts we empirically verified). In many cases, we can assume
that two persons who were exposed to similar information have the same
preferences. If their preferences diverge, we
look for a difference on their background  knowledge and motives.
What constitutes a basis of preference is beyond the scope of this
paper but labeling on the basis of  criteria as the above
gives preference a social dimension, which in turns makes preference
a basis of reasoning.

% Classical connectives and rules alone are not sufficient to represent
% preference. If that was the case then we could decide preference on
% syntactic representation alone. For example, if  two persons  have
% an identical set $K$ of
% beliefs (deductively closed set of classical sentences) then this set
%  might become distinct by
% preferring a different explanation for a new belief acquire. In other
% words, their revision functions $*_1$ and $*_2$ may be distinct so
% that, for a given sentence $\ga\not\in K$, we have
% $K*_1\ga\not=K*_2\ga$.  It is
% exactly this dynamic dimension of preference, related to abduction,
% belief revision and, in general,  belief state transformation, that
% cannot be captured by classical reasoning.

What is the logic of preference? A  simple but crucial first step has
been made by Shoham (\cite{SHO87},\cite{SHO88})
with the introduction of preferential models. Preferential models are
models equipped with a (non-reflexive, transitive)
preference ordering. Models of this sort are not,
strictly speaking, new as they can be reduced to Kripke
models or some other labeled order or relation. What is original about
them is the nature of the preference relation. This relation seeks
to maximize some  function. To make this point clearer, let
us suppose we have some box emitting binary streams, it has emitted
$000$ until now, and we want to order two binary streams $0000$ and $0001$
according to our preference for its future behavior. Our first
impulse would be to rank them equally, as both are possible.
This is what we would do if we knew nothing about the box. However,
some background information  might make us choose one over
the other, for example $0000$. In both cases, (conditional) probability
would prevail. On the other hand, if some profit
is to be made by choosing the less probable  $0001$ then again our ordering
would be a biased one. This preference would seek to maximize utility.

The above discussion points implicitly to conditional information and
therefore to nonmonotonic inference
defined through preferential models. Indeed, what Shoham did
is, by fixing  a preferential model,
to define: $\ga$  {\em preferentially entails\/} $\gb$
iff $\gb$ holds on all minimal models of $\ga$ under the preference
relation. Preferential entailment is  {\em nonmonotonic\/} as
minimal models of $\ga\land\gc$ might differ from those of $\ga$.
The preferential model approach to nonmonotonicity is a semantical
oasis in the overridden world of syntactic nonmonotonic formalisms. It
should be pointed out, however, that preferential models have their
roots in McCarthy's Circumscription (\cite{MCA80}) as the latter is a
syntactic
formalism of selecting the minimal models in a relation that prefers
predicates with a smaller extension.

The second important step was made by subsequent work of Kraus,
Lehmann, and
Magidor (\cite{KLM90}) when they showed that preferential entailment
on models whose preferential relation
satisfies the additional second order property of {\em smoothness\/} or
{\em stopperedness\/} is characterized by the  the system $\bf P$
(see Table~\ref{table:rules}),
\begin{table}[!t]
\caption{System {\bf P}}
\label{table:rules}
\begin{center}
\fbox{
%\begin{minipage}{4.5in}
%\begin{minipage}{5in}  % for fullpage
{\small
\begin{tabular}{cl}
$\displaystyle \frac{\ga \vd \gb}{\ga\ev\gb}$ & {\footnotesize (Supraclassicality)}
 \vspace{.08in} \\

 $\displaystyle \frac{\ga\vd\gb \quad \gb\vd\gc \quad\ga\ev\gc}{\gb\ev\gc}$ &
{\footnotesize (Left Logical Equivalence)}\vspace{.08in} \\

$\displaystyle \frac{\ga\ev\gb \qquad \gb\vd\gc}{\ga\ev\gc}$ &
{\footnotesize(Right Weakening)} \vspace{.08in} \\

 $\displaystyle \frac{\ga\ev\gb \qquad
\ga\ev\gc}{\ga\ev\gb\land\gc}$ & {\footnotesize(And)} \vspace{.08in}\\

$\displaystyle \frac{\ga\ev\gb \qquad
\ga\land\gb\ev\gc}{\ga\ev\gc}$ & {\footnotesize(Cut)} \vspace{.08in}\\

$\displaystyle \frac{\ga\ev\gb \qquad
\ga\ev\gc}{\ga\land\gb\ev\gc}$ & {\footnotesize(Cautious
Monotonicity)}\vspace{.08in}\\

$\displaystyle \frac{\ga\ev\gc \qquad
\gb\ev\gc}{\ga\lor\gb\ev\gc}$ & {\footnotesize(Or)} \vspace{.08in} \\

  $\displaystyle
\frac{\ga\lor\gb\ev\ga \qquad
\gb\lor\gc\ev\gb}{\ga\lor\gc\ev\ga}$ &  {\footnotesize(Weak Transitivity)}
\vspace{.08in}\\

\end{tabular}
}
%\end{minipage}
}
\end{center}
\end{table}
where $\ga\ev\gb$ means $\ga$ preferentially entails $\gb$.
This result made a connection between the preferential model approach
and work on (sceptical) nonmonotonic consequence
operators introduced by Gabbay (\cite{GAB85}) and studied by Makinson
(\cite{MAK89}).
System $\bf P$ is a simple yet powerful
 sequent-like consequence relation that has been recognized
(\cite{KLM90},\cite{MAK94}) as the strongest basis for
nonmonotonic inference.  Any system
stronger than $\bf P$ is bound to be non-Horn and therefore loose some of
its proof-theoretic content. However, apart from greatly diverging from
the theory of (monotonic) logical consequence, preferential entailment
 has the additional defect of the inability of expressing credulous
 nonmonotonic inference, that is, to express extensions.

 The purpose of this paper is to introduce  a binary relation among sentences, called {\em partial entrenchment\/}, that has the  feature of being {\em monotonic\/}
and  {\em express extensions\/} and show that any class satisfying system $\bf P$ can be
generated as the intersection of those extensions. The subclass of partial entrenchments consisting of
total preorders is G\"ardenfors and Makinson's {\em expectation
orderings\/} which  characterize expectation
inference (\cite{GM94}) and Lehmann and Magidor's rational inference
(\cite{LM92}).
Restricting the class of expectation orderings with properties
parameterized by theories one gets {\em epistemic entrenchment\/}, a
well known class of linear preorders of sentences characterizing the
AGM postulates for belief revision (\cite{AGM85}). A further generalization of partial entrenchment led to a uniform characterization of all
nonmonotonic inference relations (\cite{KG97b}).

The plan of this paper is as follows.
In Section~\ref{partial-entrenchment}, we shall introduce partial entrenchment,
explain its function and compare its features with other approaches. In Section~\ref{maxiconsistent-inference}, we  define a nonmonotonic consequence relation based
on partial entrenchment called {\em maxiconsistent\/} inference and prove some
of its properties.
Maxiconsistent consequence satisfies the
properties of system $\bf P$ and, in
Section~\ref{preferential-to-entrenchment}, we show that once we restrict
the class of partial entrenchment to an appropriate subclass  we get
a bijective correspondence.

\section{Partial Entrenchment}
\label{partial-entrenchment}

In this paper, we will not give a semantic account of entrenchment
relations but a procedural one. We will now proceed with the
formal definition of partial entrenchment. We will use a propositional
language of atomic
variables, denoted by Greek lower case letters $\ga$, $\gb$, $\gc$,
etc., and closed under the usual
propositional connectives $\neg$ (negation), $\lor$ (disjunction),
$\land$ (conjunction), and $\lto$ (implication). Entrenchment relations
assume an underlying logic. We will use  classical propositional
consequence denoted with $\vd$. Such a choice is almost dictated by the
choice of connectives and the theory we will develop but, in addition, our
intention is to build non-classical reasoning on top of a classical
one. This has the advantage of making our choices simpler and clearer.
The set of  consequences of a set of sentences $X$ under $\vd$, will
be denoted by $\Cn(X)$ and  we will write $\Cn(\ga)$ and $\Cn(X,\ga)$
for $\Cn(\{\ga\})$ and $\Cn(X,\{\ga\})$, respectively.

\begin{defn}
A binary relation $\leq$ on $\cl$ is called {\em a partial entrenchment}  when
it satisfies the following properties:
\medskip

\begin{tabular}{ll}
1. if $\ga\leq\gb$ and $\gb\leq\gc$, then $\ga\leq\gc$, & (Transitivity) \\
2. if $\ga\vd\gb$, then $\ga\leq\gb$, and & (Dominance) \\
3. if $\gc\leq\ga$ and $\gc\leq\gb$ then $\gc\leq\ga\land\gb$. &
                                                         (Conjunction)
\end{tabular}

We write $\ga<\gb$ for $\ga\leq\gb$ but $\gb\not\leq\ga$.
\end{defn}

Partial entrenchment relations can be read as rules for extending
theories. The
meaning of $\ga\leq\gb$, where $\leq$ denotes the partial entrenchment
is:
\begin{quote}
 $\ga$ can extend our theory  {\em provided\/} we first extend it
 with  $\gb$.
\end{quote}
So entrenchment encodes {\em constraints\/} on theory extensions.
Therefore,  entrenchment is a priority mechanism for {\em building\/} extensions:
we shall consider only extensions that satisfy the entrenchment rules.
The larger the extension the better.
The reader can easily verify that our reading of partial entrenchment
satisfies the above properties.

Partial entrenchment can also be expressed
as a consequence relation
that extends classical logic.
The main point here is that partial entrenchment  respects
neither disjunction nor negation.

We shall now describe informally how entrenchment gives rise to a nonmonotonic consequence relation, i.e., a conditional of the form $\ga\ev\gb$.
This paper is taking a different approach on entrenchment by defining inference {\em directly} on an entrenchment preorder. Here, the entrenchment relation becomes the primary basic notion and nonmonotonic inference takes a secondary higher-order place much like any consequence relation given some underlying proof theoretic mechanism. Entrenchment is not a proof mechanism as it lacks truth functionality but is essentially a priority preorder encoding our preferences.
Inference can be roughly described as follows:
\begin{quote}
In order to evaluate a conditional $\ga\ev\gb$ drop all sentences that could imply $\neg\ga$. What remains are the sentences compatible with $\ga$. Form all maximal consistent subsets and consider their intersection. Then $\ga\ev\gb$ holds if $\ga$ together with this set implies $\gb$.

\end{quote}

Similar proposals for evaluating conditionals have a long history in the philosophical logic literature going back to Lewis (\cite{LEW73}) (see \cite{FHL94} for a relevant discussion). Entrenchment is the mechanism for keeping track of this compatibility relation. A sentence is compatible with $\ga$ (we use {\em coherent} in Definition~\ref{definition:maxiconsistent-inference}) if it is not less than $\neg\ga$. This is also the main idea of G\"ardenfors and Makinson. The novelty of our work is that we consider partial preorders and show that the same way of evaluating conditionals still applies, giving rise to preferential inference. As partial preorders give a multitude of possible maximal compatible sets we consider their intersection, that is a 'sceptical' sort of inference.

%\begin{figure}[!ht]
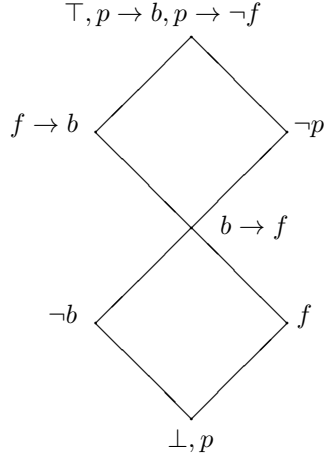
\begin{figure}
\centering
\setlength{\unitlength}{3947sp}%
\begin{picture}(1812,2880)(941,-2797)
\thinlines
\put(2101,-61){\line(-1,-1){600}}
\put(2101,-61){\circle*{2}}
\put(1501,-661){\line( 1,-1){600}}
\put(1501,-661){\circle*{2}}
\put(2101,-1261){\line(-1,-1){600}}
\put(2101,-1261){\circle*{2}}
\put(1501,-1861){\line( 1,-1){600}}
\put(1501,-1861){\circle*{2}}
\put(2101,-2461){\line( 1, 1){600}}
\put(2101,-2461){\circle*{2}}
\put(2701,-1861){\line(-1, 1){600}}
\put(2701,-1861){\circle*{2}}
\put(2101,-1261){\line( 1, 1){600}}
\put(2701,-661){\circle*{2}}
\put(2701,-661){\line(-1, 1){600}}
\put(1301,41){$\top, p \lto b, p \lto \neg f$}
\put(2741,-661){$\neg p$}
\put(961,-661){$f \lto b$}
\put(2281,-1301){$b \lto f$}
\put(2761,-1861){$f$}
\put(1201,-1861){$\neg b$}
\put(1951,-2641){$\bot, p$}
\end{picture}
\caption[]{A (transitive) entrenchment relation.}{\label{fig:tree}}
\end{figure}

Consider the following simple example. In Figure~\ref{fig:tree},
a path upwards from $\ga$ to $\gb$ indicates that $\ga\leq\gb$, where
$\leq$
denotes the entrenchment relation.
The partial entrenchment of Figure~\ref{fig:tree} says, for example, that
$\bot$ is less entrenched than all formulas, $f$ is less entrenched than
$\neg p$,
$b\lto f$ and
$f \lto b$, while $f\lto b$ is less entrenched than  $p\lto b$, $p\lto
\neg f$ and $\top$.

For instance, let us assume
$p$ and suppose we want to extend the classical theory of $p$, $\Cn(p)$
to a consistent theory. We can add any sentence to it, provided we do not add
$\neg p$ or any sentence implying $\neg p$. However, our entrenchment
example says that apart from  $\neg p$ and any sentence implying $\neg p$ we
should exclude any sentence {\em less\/} than $\neg p$ in the entrenchment
relation. We shall see that the definition of entrenchment will ensure us that
$\neg p$ and all sentences stronger than $\neg p$ are less than $\neg
p$ in the entrenchment relation. So we can use  the entrenchment
relation alone and exclude all  sentences less than $\neg p$.
So we are left with
$\{p\lto b, p\lto\neg f, f\lto b\}$. We can add those to $\Cn (p)$ to
form the extension $\Cn(p,b,\neg f)$.

Now, let us assume nothing but true sentences and see how we can extend
$\Cn (\emptyset)$. As before, we should only exclude
formulas less or equal to $\bot$. In this case, we cannot consider
together all
sentences that are {\em not\/} less or equal to $\bot$, because this set of
sentences is inconsistent. However, we can choose consistent subsets
 from this set. We must only take care that such sets are
{\em upper closed\/} so that they obey the entrenchment relation constraints.
Further, we want to add as many sentences as possible so these sets
must be maximal.
There are two such upper-closed
maximal consistent sets of sentences: one contains $\neg b$ and $f\lto b$ and
the other $f$ and $f\lto b$. Adding those to $\Cn (\emptyset)$, we can form two
extensions:
$\Cn(\neg b,\neg f,\neg p)$ and $\Cn(b, f, \neg p)$.
Therefore, it is possible to have more than one
alternative for extending the theory of our assumptions leading to the
well-known phenomenon of multiple extensions.

Considering non-truth functional orderings of sentences while
respecting conjunction is rather an old idea,
going back to Schackle (\cite{SHAC61}), and used in different disguises
in works of
Levi (\cite{LEV66}), Cohen (\cite{COH73}), Shafer (\cite{SHAF76}),
Zadeh (\cite{ZAD78}), Spohn (\cite{SPO87}), and Dubois and Prade
(\cite{DUP91}). The
above authors use an ordering of sentences satisfying the partial
entrenchment properties. However, they impose an additional constraint:
\begin{quote}
for all $\ga,\gb\in\cl$, either $\ga\leq\gb$ or $\gb\leq\ga$.
\qquad (Connectivity)
\end{quote}
A partial entrenchment satisfying connectivity will be called \emph{connected}.
The important
contribution of G\"ardenfors and
Makinson was to show that such connected preorders characterize exactly
(not only define) expectation inference. Subsequently, the author showed
that these orderings characterize also Lehmann and Magidor's rational
inference in~\cite{KG96e}. The main contribution of this paper is showing that
dropping the connectivity condition, the resulting class of orderings,
that is, the class of partial entrenchment defined above, gives rise to preferential inference
as a sceptical form of nonmonotonic inference.

 Lindstr\"om and Rabinowicz (\cite{LR91}) were the first to propose dropping
connectivity from the G\"ardenfors-Makinson connected entrenchment. Their {\em epistemic    entrenchment orderings\/}
form a
subclass of partial entrenchment by satisfying additional
 postulates related to a fixed theory and were used for describing a relational belief revision system. Their approach is slightly different
 to ours as they
require an overall consistent entrenchment. However, a common central idea of
both approaches  is that such relations point to more than one extension.

The linear preorder that G\"ardenfors and Makinson introduced by the name of epistemic entrenchment had apart from connectivity two other important features: Transitivity and Dominance.  Transitivity shows that we deal with a simple notion of transitive preference while Dominance shows that more specific sentences should be prefered over more general ones.  These properties are the basic characteristics of entrenchment and form also a part of the definition of partial entrenchment.

There are at least two other previous attempts of characterizing nonmonotonic inference through some ordering of sentences. These are Michael Freund's {\em preferential orderings} (\cite{FRE93}) and Hans Rott's {\em generalized epistemic entrenchments} (\cite{ROT92}). Both have a similar approach giving a correspondence with nonmonotonic consequence relations\footnote{Strictly speaking, Rott is characterizing weaker than rational non-Horn belief contraction systems.}.  Both build on a syntactic condition that translates  rational consequence relations to preorders. Hans Rott is using the G\"ardenfors and Makinson condition
on belief contractions
while Freund is using the Kraus, Lehman and  Magidor condition ($\ga\leq\gb$ iff $\ga\ev\ga\lor\neg\gb$). In order to generate a preferential inference relation they consider a translation of a connected entrenchment: Freund is using the contrapositive (page 236 in \cite{FRE93}) and Rott the complement of the inverse. Then they relax properties of the translated entrenchment.  However this approach leads to preorders that if, they are translated back to entrenchment would fail either Dominance in Freund's case (property $P1$, page 237 in \cite{FRE93}) or Transitivity in Rott's case (SEE1, page 52 in \cite{ROT92}).

This loss of these properties is not however the main difference between the work presented here and those proposals. Those proposals insist on generating consequence relation in a deterministic way given a preferential ordering. In a partial setting, preference gives rise to more than one alternative, that is. a multitude of most preferred possible situations and the process of inferring statements becomes nondeterministic.

%%%%%%%%%%%%%%%%%%%%%%%%%%%%%%%%%%%%%%%%%%%%%%%%%%%%%%%%%%%%%%%%%%%%%%%%%%%%%%%%

\section{Maxiconsistent Inference}
\label{maxiconsistent-inference}
We shall now proceed in describing nonmonotonic inference through partial
entrenchment. In defining inference, we shall make heavy use of
negation, or better, of consistency. This is a very important point
often overlooked by previous works on entrenchment. This is the only
place where entrenchment makes effective use of the underlying logic, in our
case, classical logic. Inference, as illustrated in the above example,
consists of two steps. First, we exclude all sentences less than the
negation of our assumption. Second, we choose maximal, upper-closed,
consistent, deductively closed sets of sentences that form our extensions.
Adding to those extensions the classical theory of our assumptions and
closing under intersection yields the nonmonotonic theory of our
assumptions. This procedure only makes sense for a finite set of
assumptions, as negation plays a central role in its definition,
so the resulting nonmonotonic consequence relation is a
subset of $\cl\times\cl$.

A partial entrenchment relation is clearly a partial preorder. A subset
$F$ of $\cl$ will be called {\em upper-closed\/}  iff $\ga\in F$
and $\ga\leq\gb$ implies $\gb\in F$. A subset
$F$ of $\cl$ will be called {\em closed under conjunction\/}  iff  $\ga,\gb\in F$
implies $\ga\land\gb\in F$. An upper-closed, closed under conjunction,
proper subset $F$ of $\cl$, is a {\em filter\/}. A filter $F$ of the partial
entrenchment is also a filter of the Boolean-Lindenbaum algebra of $\vd$
and, therefore, deductively closed, that is, $\Cn(F)=F$. The converse
is not true. A deductively closed $F$ set might fail to be a partial
entrenchment filter.  However, the  upper-closure
$\uparrow\! F$ of $F$ is the least filter containing $F$. This fact is a
consequence of Dominance and Conjunction. Principal upper closed sets
are filters and deductively closed, that is, $\uparrow\!\ga=\Cn(\ga)$.

Given a partial entrenchment, we shall denote its set of filters with
$\cf$. The space $\langle \cf, \sbs\rangle$ is itself a complete semilattice
with intersection as meet. It has also directed joins because if two
filters are included in a third then the intersection of all filters
containing their union is again a filter. This kind of partial order is
often called a dcpo.

\begin{defn}
\label{definition:maxiconsistent-inference}
Let $\leq $ be a partial  entrenchment.
The set of {\em coherent sentences\/}
for a sentence $\ga\in\cl$ is the set
$$ \Coh(\ga)=\{\gb\mid \gb\not\leq\neg\ga\}.$$
The {\em base\/} of  $\ga$ is the set
$$\cf(\ga)\quad=\quad\{ F \mid F \in \cf,\ F\sbs \Coh(a)\}.$$
The {\em maximal base\/} of $\ga$ is the set
$$\cf_{\max}(\ga)=\{ F \mid F \in\cf(\ga),\ \hbox{and if}\
F'\in\cf(\ga)
\ \hbox{with}\ F\sbs F' \ \hbox{then} \ F=F'\}.
$$

The
{\em extension set\/} of $\ga$ is the set
$$e(\ga)=\{\Cn(F,\ga)\mid F\in\cf_{\max}(\ga)\}.$$
The {\em sceptical extension\/} of $\ga$ is the set
$$E(\ga)=\bigcap e(\ga),$$ and
now define
$$\ga\sev\gb\qquad\hbox{iff}\qquad \gb\in E(\ga),$$ and say that {\em $\ga$
maxiconsistently infers
$\gb$ in the partial entrenchment} $\leq$.
\footnote{Filters have been employed by Lindstr\"om and Rabinowicz for
defining multiple revision outcomes. In~\cite{LR91}, our $\cf(\ga)$ and
$\cf_{\max}(\ga)$ are called {\em fallbacks\/} and
{\em maximal fallbacks\/} of $\ga$.}

\end{defn}

Note that
$$ F\in\cf(\ga)\quad\hbox{iff}\quad\neg\ga\not\in F\quad\hbox{iff}\quad
\neg\ga\not\in\Cn(F).$$
Unless $\leq$ equals $\cl\times\cl$, i.e. the inconsistent
ordering, $\cf$ is non-empty. As a corollary of Zorn's lemma, every filter
not containing $\neg\ga$ is included in
an element of $\cf_{\max}(\ga)$. Therefore, if $\cf(\ga)$ is non-empty then
$\cf_{\max}(\ga)$ is non-empty. On the other hand, $\cf(\ga)$ can be
empty, even though $\leq$ is not inconsistent. This can only happen if
$\gb\leq\neg\ga$, for all $\gb\in\cl$. In this case, we have that
$\ga\ev\bot$. In fact we have the following
$$e(\ga)=\emptyset \quad\hbox{iff}\quad \bigcap e(\ga)=\cl
  \quad\hbox{iff}\quad \ga\ev\bot \quad\hbox{iff}\quad
  \top\leq\neg\ga.$$

The following properties of bases will be useful in the subsequent proofs.

\begin{lem}\label{lem:base-properties}
For all $\ga,\gb\in\cl$ we have:
\begin{enumerate}
\item
if $\ga\vd\gb$ then $\cf(\ga)\sbs\cf(\gb)$,\label{infer}
\item
$\cf(\ga\lor\gb)=\cf(\ga)\cup\cf(\gb)$,\label{disjunction}
\item
\label{max-disjunction}
$\cf_{\max}(\ga\lor\gb)=(\cf_{\max}(\ga)\cap\cf_{\max}(\gb))\cup
(\cf_{\max}(\ga)\backslash\cf(\gb))\cup(\cf_{\max}(\gb)\backslash\cf(\ga))$,
\item
$\cf(\ga\land\gb)\sbs\cf(\ga)\cap\cf(\gb)$,
\label{conjunction}
\item
if $\ga\ev\gb$ then $\cf(\ga)=\cf(\ga\land\gb)$,
\label{ev-conjunction}
\item
 $\ga\notev\gb$ if and only if $\cf_{\max}(\ga)\cap\cf_{\max}(\ga\land\gb)\not=\emptyset$.
\label{rational}
\end{enumerate}

\end{lem}

\begin{pf}
We have $\neg\gb\vd\neg\ga$ which implies $\neg\gb\leq\neg\ga$. This shows that
if $F$ is a filter and $\neg\ga\not\in F$ then $\neg\gb\not\in F$ and we conclude
Part~(\ref{infer}).

For the right to left direction of Part~(\ref{disjunction}), use
Part~(\ref{infer}) to show that $\cf(\ga)\cup\cf(\gb)\sbs\cf(\ga\lor\gb)$. For the
other direction, observe that if
$\neg\ga\land\neg\gb\not\in F$ then either $\neg\ga\not\in F$ or
$\neg\gb\not\in F$,
since $F$ is closed under conjunctions. Hence $F\in \cf(\ga)\cup\cf(\gb)$ and we
conclude Part~(\ref{disjunction}).

For the left to right inclusion of Part~(\ref{max-disjunction}), assume
$F\in\cf_{\max}(\ga\lor\gb)$. Observe that $F\in \cf(\ga)$
implies $F\in \cf_{\max}(\ga)$, else there exists
$F'\in\cf_{\max}(\ga)$ such that $F\sbs F'$ and $F\not= F'$. We have
$F'\not\in \cf_{\max}(\ga\lor\gb)$, since $F\in\cf_{\max}(\ga\lor\gb)$.
So $F'\not\in\cf(\ga)$, by Part~\ref{infer}, a contradiction.

For the other inclusion, assume
$$F\in(\cf_{\max}(\ga)\cap\cf_{\max}(\gb))\cup
(\cf_{\max}(\ga)\backslash\cf(\gb))\cup(\cf_{\max}(\gb)\backslash\cf(\ga)).$$
Let $F'\in\cf(\ga\lor\gb)$ with $F\sbs F'$. We have either
$F'\in\cf_{\max}(\ga)$ or $F'\in\cf_{\max}(\gb)$, by
Part~(\ref{disjunction}). In the first case, we have $F\in
\cf(\ga)$, as $\cf(\ga)$ is lower-closed, and this can only happen if
$F\in(\cf_{\max}(\ga)\cap\cf_{\max}(\gb))\cup
(\cf_{\max}(\ga)\backslash\cf(\gb))$. So $F\in\cf_{\max}(\ga)$ and
$F=F'$. The other case is similar and, therefore,
$F\in\cf_{\max}(\ga\lor\gb)$.

Part~(\ref{conjunction}) is a straightforward corollary
of Part~(\ref{infer}).

Now, we turn to Part~(\ref{ev-conjunction}). By Part~(\ref{conjunction}), we have
$\cf(\ga\land\gb)\sbs\cf(\ga)$. If $\cf(\ga)=\emptyset$, we are done.
Suppose that $\cf(\ga)\not=\emptyset$, and let
$F\in \cf(\ga)$. Further, let $F'\in \cf_{\max}(\ga)$ such
that $F\sbs F'$. By our hypothesis, we have $F',\ga\vd\gb$. Also, we have
$\ga\lto\neg\gb\not\in F'$, since otherwise $\neg\ga\in F'$.
Therefore $\neg\ga\lor\neg\gb\not\in F$. Hence
$F\in\cf(\ga\land\gb)$.

For Part~\ref{rational}, suppose that $\ga\notev\gb$ then there exists $F\in\cf_{\max}(\ga)$ such that $\neg\ga\lor\gb\not\in F$. Therefore, $F\in\cf(\ga\land\gb)$ and, since $\cf(\ga\land\gb)\sbs\cf(\ga)$, $F\in\cf_{\max}(\ga\land\gb)$. The other direction is similar.
\qed\end{pf}

 It is worth noting that from the algebra of sentences we moved to the
 algebra of theories and onto the algebra of the powerset of theories. The
 last algebra is of considerable interest as is the algebra pertaining to
 nonmonotonic inference. For example, we could dispense
 with  maximal filters and study directly the lattice of the powerset
 of $\cf$. Our intention, however, is to
 introduce as little theory overhead as possible.

 We now have everything we need for characterizing preferential
 inference. However, we should first verify our claim that maxiconsistent inference
 is a preferential one.

  \begin{thm}\label{thm:p-soundness}
   Given a partial entrenchment $\leq$, the consequence relation $\sev$
   satisfies the system $\bf P$ rules.
  \end{thm}

  \begin{pf}
   We  verify the following  list of rules: Supraclassicality,
Left Logical Equivalence, Right Weakening, And, Cut, Cautious Monotony,
 and
Or. \footnote{Cut is redundant, see~\cite{KLM90}.}

For Supraclassicality, suppose that $\ga\vd\gc$ then $F,\ga\vd\gc$, for all $F\in
\cf_{\max}(\ga)$.

For Left Logical Equivalence, suppose that $\ga\vd\gb$ and $\gb\vd\ga$. By
Lemma~\ref{lem:base-properties}(\ref{infer}) we have that
$\cf_{\max}(\ga)=\cf_{\max}(\gb)$. So, for every filter
$F\in\cf_{\max}(\ga)=\cf_{\max}(\gb)$, if $F,\ga\vd\gc$ then $F,\gb\vd\gc$.

For And, suppose that $F,\ga\vd\gb$ and $F,\ga\vd\gc$, for all
$F\in\cf_{\max}(\ga)$.
Then $F,\ga\vd\gb\land\gc$.

For Right Weakening, Suppose that for all $F\in\cf_{\max}(\ga)$ we have
$F,\ga\vd\gb$
and $\gb\vd\gc$. Then by (classical) Cut we get $F,\ga\vd\gc$.

For Cut, suppose that
$\ga\ev\gb$ and
$\ga\land\gb\ev\gc$. Suppose that $F \in \cf_{\max}(\ga)$ then $F,a\vd\gb$.
By Lemma~\ref{lem:base-properties}(\ref{ev-conjunction}), we have
$\cf_{\max}(\ga\land\gb)$ and, therefore, $F,\ga\land\gb\vd\gc$.
 By (classical) Cut, we have $F,\ga\vd\gc$. Hence $\ga\ev\gc$.

For Cautious Monotony, suppose that $\ga\ev\gb$ and $\ga\ev\gc$, and let
$F\in\cf_{\max}(\ga\land\gb)$. By
Lemma~\ref{lem:base-properties}(\ref{ev-conjunction}), we have $F\in
\cf_{\max}(\ga)$. Thus $F,\ga\vd\gc$, and therefore $F,\ga\land\gb\vd\gc$.
Hence
$\ga\land\gb\ev\gc$.

For Or, suppose that $\ga\ev\gc$ and $\gb\ev\gc$, and let
$F\in\cf_{\max}(\ga\lor\gb)$. By
Lemma~\ref{lem:base-properties}(\ref{conjunction}), there are three cases to
consider: either (i) $F\in\cf_{\max}(\ga)\cap\cf_{\max}(\gb)$, or (ii)
$F\in\cf_{\max}(\ga)$ with $F\not\in\cf(\gb)$,
or (iii) $F\in\cf_{\max}(\gb)$ with
$F\not\in\cf(\ga)$. In case (i), we have $F,\ga\vd\gc$ and
$F,\gb\vd\gc$,
so $F,\ga\lor\gb\vd\gc$. In  case (ii),  we have
$F,\gb\vd\bot$, and therefore $F,\gb\vd\gc$. Again $F,\ga\lor\gb\vd\gc$,
as above. Case (iii) is similar.
\qed
\end{pf}

Given the above results we can now give a simple translation of the property of Rational Monotonicity.

\begin{cor}
\label{lem:rational}
Let $\leq$ be a partial entrenchment. Then $\sev$ satisfies
$$
\frac{\ga\notsev\neg\gb \qquad
\ga\sev\gc}{\ga\land\gb\sev\gc}\qquad\hbox{(Rational Monotonicity)}$$
if and only if
\begin{center}
$\cf_{\max}(\ga)\cap\cf_{\max}(\ga\land\gb)\not=\emptyset$ and $\ga\lto\gc\in\bigcap\cf_{\max}(\ga)$ implies
$\ga\land\gb\lto\gc\in\bigcap\cf_{\max}(\ga\land\gb)$.
\end{center}
\end{cor}

\begin{pf}
It is immediate by the definition of maxiconsistent inference and Lemma \ref{lem:base-properties}.\ref{rational}.
\qed\end{pf}

A very natural subclass of partial entrenchments is the original class of connected entrenchments introduced by G\"ardenfors and Makinson (\cite{GM88},\cite{GM94}). This class was shown to be equivalent to the class of rational nonmonotonic consequence relations (\cite{KG96e}) under the following translation:

\begin{quote}

\begin{tabular}{rcl}
 $a\gmev\gb$ & iff & either $\gb\leq\neg\ga$, for all $\gb\in\cl$, \\
& & or there is a $\gb\in\cl$ such that $\{\gb \mid \ga< \gb\} \vd \ga\lto\gc $.
\end{tabular}

\end{quote}

 It is easy to see, by Definition~\ref{definition:maxiconsistent-inference}, that the above way through which a connected entrenchment gives rise to a nonmonotonic consequence relation is a special instance of the definition of maxiconsistent inference, that is, $\sev=\gmev$. Now we can give an alternative proof of the fact that connected entrenchments give rise to rational nonmonotonic consequence relations by showing that a connected entrenchment satisfies the property of Lemma~\ref{lem:rational}. In fact, it satisfies a much stronger property as the following lemma shows.

\begin{lem}
 If $\leq$ is a connected entrenchment then
\begin{center}
$\cf_{\max}(\ga)\cap\cf_{\max}(\ga\land\gb)\not=\emptyset$ implies $\cf_{\max}(\ga\land\gb)\sbs\cf_{\max}(\ga)$.
\end{center}
\end{lem}

\begin{pf}
This is immediate because if $\leq$ is connected then $\cf_{\max}(\ga)$ is either a singleton or empty for all $\ga\in\cl$.
\qed\end{pf}

\begin{cor}[\cite{GM94},\cite{KG96e}]
If $\leq$ is a connected entrenchment then $\sev$ is a rational inference relation.
\end{cor}

In the next section we shall exhibit a class of non-connected entrenchment relations that satisfy the property of Lemma~\ref{lem:rational}, and therefore give rise to rational inference relations.
We leave open the question whether there is a simple first-order property of $\leq$ that relaxes connectivity and still implies the property of Lemma~\ref{lem:rational}.
The above corollary shows why maxiconsistent inference makes partial entrenchment a generalization of the G\"ardenfors and Makinson original notion of entrenchment.
It is well known that connected entrenchments  not only give rise to rational inference but they are in bijective correspondence as well. Given a rational inference relation $\ev$ one can construct a connected entrenchment $\leq$ with $\ev=\sev$ using the translation below
\begin{quote}
$\ga\leq\gb$ \quad iff \quad either $\vd\ga\land\gb$ or $\neg(\ga\land\gb)\notev\ga$,
\end{quote}
proposed in \cite{KG96e} which is a slightly modified version of the one proposed by G\"ardenfors-Makinson for expectation inference relations (see \cite{GM94}). In the case of partial entrenchment relations the above translation no longer works. In the next section, an alternative way to generate  entrenchment given a preferential inference relation will be presented.
\section{Reducing Preferential Inference to Partial Entrenchment}
\label{preferential-to-entrenchment}

In this section, we  show that every preferential consequence
relation can be expressed as a maxiconsistent inference of a partial
entrenchment.
The class of maxiconsistent inference relations is much wider than
that of preferential inference. Maxiconsistent inference expresses
sceptical nonmonotonic consequence by an intersection of possible
extensions. Therefore, we can construct two different partial
entrenchments assigning different sets of extensions for the same
assumptions while still  agreeing on the intersection of the extensions.

Given a preferential inference relation, we will construct a partial
entrenchment with the same maxiconsistent inference. This construction
will be canonical, in the sense that one can safely identify a
preferential inference relation with the partial entrenchment
constructed. The main idea is to construct a partial entrenchment with
all possible extensions of the sceptical extension.
 This way their intersection will also
 provide the sceptical extension. It turns out that such partial
entrenchments can be described syntactically by adding the following
rule to Dominance, Transitivity, and Conjunction. For all
$\ga,\gb,\gc\in\cl$
\begin{quote}
if $\ga\lto\gb\leq\neg\ga$ and  $\ga\lto\gc\leq\neg\ga$
then $\ga\lto(\gb\lor\gc)\leq\neg\ga$.\quad (Weak Disjunction)
\end{quote}
A partial entrenchment satisfying Weak Disjunction is called {\em
weakly disjunctive\/}. The class of weakly disjunctive
partial entrenchment
is properly contained
in that of partial entrenchments as the following simple counterexample
shows.

\begin{exmp}
Let $D=\{\phi,\phi\lor\psi\lor\chi\}$, and define an ordering as follows
$$\ga\leq\gb\qquad  \hbox{iff}  \qquad B\vd\ga\ \hbox{implies} \ B\vd\gb,\
\hbox{for all} \ B\sbs D.$$
The preorder $\leq$ is a partial entrenchment.
 However, it is not
weakly disjunctive, for $\phi\lor\psi\leq\phi$ and $\phi\lor\chi\leq\phi$ but
$\phi\lor\psi\lor\chi\leq\phi$.
\end{exmp}

 The main property of weakly disjunctive partial entrenchments
 is given in the following
proposition.

\begin{prop}\label{proposition:weak-disjunction}
Let $\leq$ be a weak disjunctive partial entrenchment. Then for all $\ga\in\cl$ and $F\in \cf_{\max}(\ga)$, either $\ga\lto\gb\in F$ or $\ga\lto\neg\gb\in F$.
\end{prop}

\begin{pf}
Fix an $\ga\in \cl$ and $F\in\cf_{\max}(\ga)$ and suppose $\ga\lto\gb\not\in F$ and $\ga\lto\neg\gb\not \in F$, towards a contradiction. As $F$ is maximal in $\cf_{\max}(\ga)$, we have $\neg\ga\in
\uparrow(F\cup\{\ga\lto\gb\})$. This implies that there exists $\gre_1\in F$ such that
$\gre_1\land(\ga\lto\gb)\leq\neg\ga$. Similarly, there exists $\gre_2\in F$ such that
$\gre_2\land(\ga\lto\neg\gb)\leq\neg\ga$. For $\gre=\gre_1\land\gre_2\in F$ we have both
$\gre\land(\ga\lto\gb)\leq\neg\ga$ and $\gre\land(\ga\lto\neg\gb)\leq\neg\ga$.
Now observe that  $(\gre\land\neg\ga)\lor(\gre\land(\ga\lto\gb))$ is classically equivalent to $\gre\land(\ga\lto\gb)$. So
$$(\gre\land\neg\ga)\lor(\gre\land(\ga\lto\gb))\leq\neg\ga.$$
Also, we have
$$(\gre\land\neg\ga)\lor(\gre\land(\ga\lto\gb))\leq\gre.$$
So, by Conjunction,
$$(\gre\land\neg\ga)\lor(\gre\land(\ga\lto\gb))\leq\gre\land\neg\ga.$$
Similarly,
$$(\gre\land\neg\ga)\lor(\gre\land(\ga\lto\neg\gb))\leq\gre\land\neg\ga.$$
Applying Weak Disjunction on the last two, we have
$$(\gre\land\neg\ga )\lor (\gre\land\ga\land\neg\gb) \lor (\gre\land (\ga\lto\gb)) \leq
\gre\land\neg\ga.$$
Therefore
$$(\gre\land\neg\ga) \lor \gre \leq \gre\land\neg\ga.$$
Since $\gre\vd (\gre\land\neg\ga )\lor \gre$, we have
$$\gre\leq  ((\gre\land\neg\ga )\lor \gre) \leq \gre\land\neg\ga \leq \neg\ga,$$
a contradiction as $\gre\in F\in\cf_{\max}(\ga)$.
\qed\end{pf}

\begin{cor}\label{cor:weak-disjunction}
Let $\leq$ be a weak disjunctive partial entrenchment. Then
$$\ga\lto\neg\gb\leq\neg\ga \qquad\hbox{iff}\qquad \ga\sev\gb.$$
\end{cor}

\begin{pf}
For the right to left direction, assume $\ga\sev\gb$ and
$\ga\lto\neg\gb\not\leq\neg\ga$. We have that $\ga\lto\gb\in F$, for all
$F\in\cf_{\max}(\ga)$, and $\ga\lto\neg\gb\in \Coh(\ga)$.
We have $\Cn(\ga\lto\neg\gb)\sbs \Coh(\ga)$.
Choose $F\in\cf_{\max}(\ga)$ such that $\Cn(\ga\lto\neg\gb)\sbs F$.
However, $F$ contains $\ga\lto\gb$ and therefore $\neg\ga\in F$,
a contradiction. Note that this direction does not use Weak
Disjunction.

For the left to right direction,
 assume $\neg\ga\lor\neg\gb\leq\neg\ga$. We must show $\ga\sev\gb$. Let $F\in\cf_{\max}(\ga)$.
By Proposition~\ref{proposition:weak-disjunction},
We have either $\ga\lto\gb\in F$ or $\ga\lto\neg\gb\in F$. We cannot have $\ga\lto\neg\gb\in F$
as $\ga\lto\neg\gb\leq\neg\ga$ so $\ga\lto\gb\in F$.
\qed
\end{pf}

We can go back and forth between a preferential inference relation and
a partial entrenchment through a syntactic translation given in the
following definition.

\begin{defn}\label{definition:ps}
Given a partial entrenchment  $\leq$ and
a nonmonotonic  consequence relation $\ev$, then define a
consequence relation $\ev'$
and a relation
$\leq'$ as follows

\begin{tabular}{rrcl}
($N$) & $\ga\ev'\gb$ & iff & $\neg\ga\lor\neg\gb\leq\neg\ga$\\
($P$) & $\ga\leq'\gb$ & iff & $\neg\ga\lor\neg\gb\ev\neg\ga$.\\
\end{tabular}

We shall  denote $\ev'$ and $\leq'$ with
$N(\leq)$ and
$P(\ev)$, respectively.
\end{defn}

Definition  $P$ is akin to a preorder defined in~\cite{KLM90} by $\ga\lor\gb\ev\ga$ (see also Makinson's comments in~\cite{MAK94}, page 78).
The maps defined in Definition~\ref{definition:ps} are
inverses of each other.

\begin{lem}\label{lemma:iso-entrenchment}
Let $\leq$ and $\ev$ be a partial entrenchment  and  a preferential
inference relation, respectively. Then
\begin{enumerate}
\item $P(N(\leq))=\leq$, and
\item $N(P(\ev))=\ev$.\label{ev-to-ev}
\end{enumerate}
\end{lem}

\begin{pf}
Let $\leq'=P(N(\leq))$. We have $\ga\leq'\gb$ iff
$\neg\ga\lor\neg\gb\ev\neg\ga$, where $\ev=N(\leq)$. Now, we have
$\neg\ga\lor\neg\gb\ev\neg\ga$ iff
$\neg(\neg\ga\lor\neg\gb)\lor\neg\neg\ga\leq\neg(\neg\ga\lor\neg\gb)$, by definition.
The latter holds iff
$(\ga\land\gb)\lor\ga\leq\ga\land\gb$  iff $\ga\leq\ga\land\gb$ , by
Dominance. Now, $\ga\leq\ga\land\gb$ implies $\ga\leq\gb$,
by Transitivity, and $\ga\leq\gb$ implies $\ga\leq\ga\land\gb$, by
Conjunction and Dominance.

Let $\ev'=N(P(\ev))$. We have $\ga\ev'\gb$ iff
$\neg\ga\lor\neg\gb\leq\neg\ga$ iff
$\neg(\neg\ga\lor\neg\gb)\lor\neg\neg\ga\ev\neg(\neg\ga\lor\neg\gb)$ iff
$\ga\ev\ga\land\gb$, by Left Logical Equivalence, iff $\ga\ev\gb$, by And, Right
Weakening and Reflexivity.
\end{pf}

Now, combining Proposition~\ref{proposition:weak-disjunction}
and Lemma~\ref{lemma:iso-entrenchment} we have the following theorem.

\begin{thm}
\label{thm:completeness}
If $\ev$ is a preferential inference
relation, then the relation $\leq$ defined by
($P$) is  a weakly disjunctive partial entrenchment relation
such that, for all
$\ga$, $\gb$ in $\cl$,
$$\ga\ev\gb \qquad\hbox{iff}\qquad
\ga\sev\gb.$$
\end{thm}

\begin{pf}
We must only show that $\leq$ is a weakly disjunctive partial
entrenchment.

For Dominance, suppose that $\ga\vd\gb$. Thus $\neg\gb\vd\neg\ga$, and so
$\neg\ga\lor\neg\gb\ev\neg\ga$. Hence $\ga\leq\gb$.

For Transitivity, suppose that $\ga\leq\gb$ and  $\gb\leq\gc$. By the definition
above, these translate to
$\neg\ga\lor\neg\gb\ev\neg\ga$ and $\neg\gb\lor\neg\gc\ev\neg\gb$, respectively.
Further, the following rule is derivable in the preferential system $\bf P$ (Lemma 5.5
in~\cite{KLM90})
$$\displaystyle \frac{\ga\lor\gb\ev\ga \qquad
\gb\lor\gc\ev\gc}{\ga\lor\gb\ev\gc}.$$
So we have $\neg\ga\lor\neg\gc\ev\neg\ga$. Hence $\ga\leq\gc$.

For Conjunction, suppose that $\gc\leq\ga$ and  $\gc\leq\gb$. We must show that
$\gc\leq\ga\land\gb$. Our assumption translates to $\neg\gc\lor\neg\ga\ev\neg\gc$ and
$\neg\gc\lor\neg\gb\ev\neg\gc$, respectively. Applying Or and Left Logical
Equivalence, we get $\neg\ga\lor\neg\gb\lor\neg\gc\ev\neg\gc$. So
$\neg(\ga\land\gb)\lor\neg\gc\ev\neg\gc$. Hence $\gc\leq\ga\land\gb$.

For Weak disjunction, suppose
$\ga\lto\gb\leq\neg\ga$ and  $\ga\lto\gc\leq\neg\ga$. These translate
to $\ga\ev\neg\gb$ and $\ga\ev\neg\gc$. By And, we have
$\ga\ev\neg\gb\land\neg\gc$.
The latter translates to $\ga\lto(\gb\lor\gc)\leq\neg\ga$ as desired.
\qed
\end{pf}

We can now give a bijective correspondence between the class of rational nonmonotonic consequence relations and weakly disjunctive partial entrenchments. It is enough to translate the property of Rational Monotonicity using $P$:
$$\displaystyle \frac{\ga\lor\neg\gb\not\leq\ga \qquad
\ga\lor\gc\leq\ga}{\ga\lor\gb\lor\gc\leq\ga\lor\gb}\qquad\hbox{(Splitting)}$$
Weakly disjunctive relations satisfying Splitting will be called {\em rational}.

\begin{cor}
\label{cor:rational-completeness}
If $\ev$ is a rational inference
relation, then the relation $\leq$ defined by
($P$) is  a rational weakly disjunctive partial entrenchment relation
such that, for all
$\ga$, $\gb$ in $\cl$,
$$\ga\ev\gb \qquad\hbox{iff}\qquad
\ga\sev\gb.$$
\end{cor}

\end{document}